\documentclass[a4paper]{jpconf}
\usepackage{graphicx}
\usepackage{amsmath, amssymb, amsgen}
\newcommand{\dd}[2]{\frac{\rmd#1}{\rmd#2}}
\newcommand{\pp}[2]{\frac{\partial#1}{\partial#2}}
\begin{document}
\title{Quantum phase transition of dynamical resistance in a mesoscopic capacitor}

\author{Yuji Hamamoto$^{1}$, Thibaut Jonckheere$^{2}$, Takeo Kato$^{3}$ and Thierry Martin$^{2,4}$}

\address{$^{1}$ Institute of Physics, University of Tsukuba, Tennodai, Tsukuba, Ibaraki 305-8571, Japan}
\address{$^{2}$ Centre de Physique Th\'eorique, Case 907 Luminy, 13288 Marseille cedex 9, France}
\address{$^{3}$ Institute of Solid State Physics, University of Tokyo, Kashiwa, Chiba 277-8581, Japan
}
\address{$^{4}$ Universit\'e de la M\'edit\'erann\'ee, 13288 Marseille cedex 9, France}

\ead{hamamoto.yuji.fp@u.tsukuba.ac.jp}

\begin{abstract}
We study theoretically dynamic response of a mesoscopic capacitor,
which consists of a quantum dot connected to an electron reservoir via a point contact
and capacitively coupled to a gate voltage.
A quantum Hall edge state with a filling factor $\nu$ is realized
in a strong magnetic field applied perpendicular to the two-dimensional electron gas.
We discuss a noise-driven quantum phase transition of the transport property of the edge state
by taking into account an ohmic bath connected to the gate voltage.
Without the noise, the charge relaxation for $\nu>1/2$
is universally quantized at $R_{q}=h/(2e^{2}\nu)$,
while for $\nu<1/2$, the system undergoes the Kosterlitz-Thouless transtion,
which drastically changes the nature of the dynamical resistance.
The phase transition is facilitated by the noisy gate voltage,
and we see that it can occur even for an integer quantum Hall edge at $\nu=1$.
When the dissipation by the noise is sufficiently small,
the quantized value of $R_{q}$ is shifted by the bath impedance.
\end{abstract}

\begin{figure}[b]
\begin{minipage}{.3\textwidth}
\includegraphics[width=\textwidth]{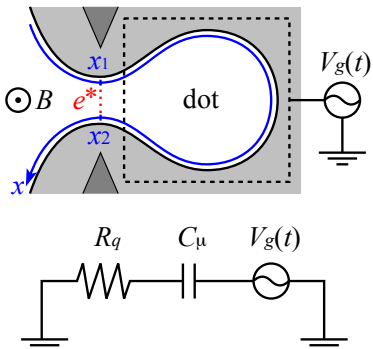}
\end{minipage}
\hfill
\begin{minipage}{.65\textwidth}
\caption{\label{fig:dot}
Schematic view of the mesoscopic capacitor and its equivalent $RC$ circuit.
The quantum dot is connected to an electron reservoir via a point contact
and capacitively coupled to a gate (shown as a dotted square) with
time-dependent voltage $V_{g}(t)$.
A magnetic field applied perpendicular to the plane
realizes a quantum Hall edge state moving along the $x$ axis.
The dotted line between $x_{1}$ and $x_{2}$ represents
quasiparticle tunneling on the point contact.
}
\end{minipage}
\end{figure}

\section{Introduction}
In the long history of the mesoscopic physics,
it is quite recent that it became possible to observe dynamic response of coherent systems in
experiments~\cite{gabelli06,gabelli07}.
Theoretically, dynamical transport properties of the mesoscopic systems has been extensively investigated
with the scattering theory~\cite{buttiker93a,buttiker93}. 
One of the most intriguing results of the theoretical efforts
is the universal quantization of charge relaxation resistance in mesoscopic
capacitors~\cite{buttiker93}.
The mesoscopic capacitors are quantum analogs of classical $RC$ circuits,
among which the simplest one (Fig.~\ref{fig:dot}) consists of a quantum dot connected
to an electron reservoir
via a quantum point contact and capacitively coupled to an ac gate voltage.
An electron entering the quantum dot
relaxes to equilibrium in the scale of charge relaxation time $\tau_{RC}$.
The dynamical resistance observed at this time scale is the so-called {\it charge relaxation resistance}.
It has been predicted by B\"uttiker {\it et al.} that the charge relaxation resistance
is quantized at half a resistance quantum $h/(2e^{2})$ irrespective of the property of the point
contact~\cite{buttiker93}.
Indeed, the universal behavior has been confirmed experimentally~\cite{gabelli06},
which has further attracted the theorists' interest in the charge relaxation
resistance~\cite{nigg06,hamamoto10,mora10}.
Recently, we have investigated theoretically the charge relaxation resistance,
treating electron interactions exactly~\cite{hamamoto10}.
We have considered a one-dimensional mesoscopic capacitor and predicted the followings;
the charging effect of the quantum dot has no influence on the quantization,
while short-range interactions described by the Luttinger parameter $K$
drastically affects the nature of the dynamical resistance.
In contrast to the universal quantization at $R_{q}=h/(2e^{2}K)$ for $K>1/2$,
the Kosterlitz-Thouless transition occurs for $K<1/2$
and as a result one can not any more observe the universal charge relaxation resistance.
In this paper, we extend our theory to discuss the quantum phase transition driven by a noisy gate voltage.
How our treatment of the charging effect is related to the conventional picture of the scattering theory
is also discussed by applying the mean field approximation.

\section{Model}
The system we have in mind is the mesoscopic capacitor
depicted in Fig~\ref{fig:dot}.
Due to the magnetic field applied perpendicular to the 2DEG,
a chiral mode called quantum Hall edge state is realized along the $x$ axis.
It has been shown by Wen~\cite{wen90} that the edge state at filling factor
$\nu$ ($1/\nu$ is an odd integer) is a chiral Luttinger liquid,
and its Hamiltonian reads
\begin{gather}
H_{0}=\frac{v}{4\pi\nu}\int_{-\infty}^{\infty}\rmd x\left(\pp{\phi}{x}\right)^{2}.
\label{eq:kinetic}
\end{gather}
Here the bosonic field $\phi$ describes the gapless edge excitations,
whose velocity is denoted by $v$.
The narrow constriction on the point contact causes intra-edge tunneling
of Laughlin quasiparticles between $x=x_{1}$ and $x=x_{2}$.
Since a Laughlin quasiparticle at $x$ is annihilated by the operator
$\propto\exp[i\phi(x)]$, the tunneling process is written as
\begin{gather}
H_{V}=V\cos[\phi(x_{1})-\phi(x_{2})],\label{eq:tunneling}
\end{gather}
where $V$ is the strength of quasiparticle tunneling
\footnote{In the argument of the cosine in Eq.~(\ref{eq:tunneling}),
we have omitted the phase that an edge mode gains along the circumference of the dot
by shifting zero of the field $\phi$.}.
Change in the dot charge $Q$ in the range of $x_{1}\le x\le x_{2}$
is strongly restricted by the long-range interaction between electrons, i.e., the charging effect.
Upon using the fact that the electron density is bosonized as
$\rho=\partial_{x}\phi/(2\pi)$, one can see that
the term describing the charging effect takes the form
\begin{gather}
H_{C}=\frac{Q^{2}}{2C}
+QV_{g}(t)\qquad\left(
Q=\frac{e}{2\pi}[\phi(x_{2})-\phi(x_{1})]\right)\label{eq:charging}
\end{gather}
where $C$ is the geometrical capacitance
essentially determined by the charging effect.
When the gate voltage is oscillating at a low frequency $\omega\ll1/\tau_{RC}$,
the admittance of the quantum $RC$ circuit can be expanded as
$G(\omega)=-i\omega C_{\mu}+\omega^{2}C_{\mu}{}^{2}R_{q}+\mathcal{O}(\omega^{3})$,
from which the charge relaxation resistance $R_{q}$ is obtained.
Note that the electrochemical capacitance defined by
$C_{\mu}\equiv-\partial\langle Q\rangle/\partial V_{g}$ in general differs from $C$,
since, e.g., the quasiparticle tunneling induces Coulomb blockade oscillation in $C_{\mu}$.
The above expansion of $G(\omega)$ is possible only when the $\tau_{RC}$ is finite;
if the phase transition occurs, $\tau_{RC}$ diverges faster than $1/T$ as $T\rightarrow0$,
so that one can not even regard $R_{q}$ as the charge relaxation resistance~\cite{hamamoto10}.

\section{Mean field theory}\label{sec:mean-field}
At a glance, our formalism described in the previous section looks slightly different from the scattering
theory~\cite{buttiker93a,buttiker93},
where the charging effect is incorporated through an effective gate voltage $U_{g}$
in the quantum dot.
Clearly, the latter treatment is justified when fluctuation of the dot charge $Q$ is negligible
enough that one can approximate the interaction of an electron with the others in the dot
with an effective potential created by mean charge $\langle Q\rangle$.
Thus one can expect that the two-capacitor obtained from the scattering theory~\cite{buttiker93}
is reproduced by applying the mean field theory to our model.
In this section, we first confirm this and compare the results of the scattering theory
and our calculation where the interaction is exactly treated.
Upon neglecting higher orders in $Q-\langle Q\rangle$ in the charging term (\ref{eq:charging}), we obtain
\begin{gather}
H_{C}{}^{\rm MF}\simeq\frac{\langle Q\rangle}{C}+QV_{g}\equiv QU_{g},\qquad
\therefore \langle Q\rangle=C(U_{g}-V_{g}),\label{eq:mean-charge1}
\end{gather}
where we have dropped constant terms.
Since an electron in the dot behaves as a free particle
subject to the effective potential $U_{g}$,
the mean charge can be also expressed using the density of states
in the dot $\rmd N/\rmd\varepsilon$ as
\begin{gather}
\langle Q\rangle=e\dd{N}{\varepsilon}(-eU_{g})\equiv -C_{q}U_{g}.
\label{eq:mean-charge2}
\end{gather}
$C_{q}$ is called quantum capacitance since it reflects
the nature of standing wave of an electron in the dot.
By eliminating $U_{g}$ from Eqs.~(\ref{eq:mean-charge1})
and (\ref{eq:mean-charge2}), we can calculate the electrochemical capacitance
\begin{gather}
C_{\mu}=-\pp{\langle Q\rangle}{V_{g}}=\frac{CC_{q}}{C+C_{q}},\qquad
\therefore\frac{1}{C_\mu}=\frac{1}{C}+\frac{1}{C_{q}}.\label{eq:series}
\end{gather}
Thus in the mean field approximation
$C_{\mu}$ is reduced to a series combination of of the classical part $C$
and the quantum part $C_{q}$,
in agreement with the results of the scattering theory
\cite{buttiker93}.
Note that $C_{q}\propto\rmd N/\rmd\varepsilon$ oscillates as $V_{g}$ is varied.
For an integer quantum Hall edge state at $\nu=1$,
one can refermionized the Hamiltonian to solve the transport problem
with the effective potential $U_{g}$ with the scattering theory.
Then the charge relaxation resistance quantized at $R_{q}=h/(2e^{2})$ will be also reproduced.

One might conjecture that the mean field assumption is valid in the limit of
weak quasiparticle tunneling $V\rightarrow0$,
where the potential that dot charge $Q$ feels is almost harmonic,
so that the charge fluctuation is suppressed upon lowering temperature.
Indeed, we can exactly derive quantized charge relaxation resistance $R_{q}=h/(2e^{2}\nu)$
up to second order in $V$~\cite{hamamoto10},
but a qualitative difference between the mean field theory and our exact result
emerges in the capacitance.
From the perturbative result~\cite{hamamoto10}, one can split $C_{\mu}$ into two parts
mimicking Eq.~(\ref{eq:series}):
\begin{gather}
\frac{1}{C_{\mu}}=\frac{1}{\gamma C}
+\frac{1}{\gamma C_{q}'}\qquad
\left(C_{q}'\equiv\frac{\nu^{2}e^{2}L}{2\pi v}\right),
\label{eq:series2}
\end{gather}
where $\gamma$ is a numerical factor oscillating as a function of $V_{g}$,
and $L\equiv x_{2}-x_{1}$ is the circumference of the dot.
For $\nu=1$, $\gamma C'_{q}$ can be identified as the quantum capacitance,
since it coincides with the small $V$ limit of $C_{q}$ obtained from the scattering theory,
while $\gamma C$ differs from the geometrical capacitance by the oscillating factor.
The latter is because the mean field approximation underestimates
the charge fluctuation ($\sim$ capacitance) due to electron tunneling between the dot and the reservoir,
which should oscillate depending on $V_{g}$.
Therefore, one can see from Eq.~(\ref{eq:series2}) that in the small $V$ limit,
$C_{\mu}$ is a series combination of $\gamma C$ and $\gamma C_{q}'$,
which reflect the particle and wave natures of an electron, respectively.

It is clear that the mean field theory breaks down at charge degenerate points
$\langle Q\rangle=(n+1/2)e$ ($n$ is an integer) for large $V$,
where the charge fluctuation can be finite down to low temperatures.
In Ref.~\cite{hamamoto10}, we have shown that
the charge relaxation resistance for $\nu>1/2$ is always quantized
at $R_{q}=h/(2e^{2}\nu)$ owing to the so-called charge-Kondo
effect~\cite{glazman90},
where two degenerate charge states $ne$ and $(n+1)e$
play the role of up and down spins of a magnetic
impurity~\footnote{Recently, an equivalent Kondo physics at $\nu=1$ has been
also discussed in Ref.~\cite{mora10}.}.
More surprisingly, charge relaxation resistance becomes ill-defined
due to the Kosterlitz-Thouless transition for $\nu<1/2$,
and the observed dynamical resistance $R_{q}$ diverges with lowering temperature.


\section{Noise-driven quantum phase transition}
\begin{figure}[t]
\begin{minipage}{.25\textwidth}
\includegraphics[width=\textwidth]{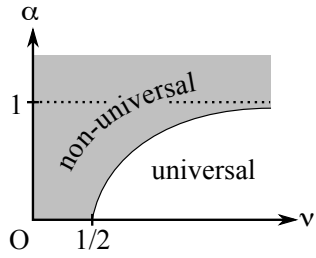}
\end{minipage}
\hfill
\begin{minipage}{.7\textwidth}
\caption{\label{fig:phase}
Phase diagram of the universal quantization of the charge relaxation resistance of the mesoscopic capacitor
with a noisy gate voltage.
$\nu$ is the filling factor, and $\alpha\equiv R_{B}/(h/e^{2})$
is the strength of dissipation, where $R_{B}$ is the bath impedance.
The charge relaxation resistance is universally quantized in the white region,
while it can undergoes a phase transition for sufficiently strong quasiparticle tunneling in the gray region.
}
\end{minipage}
\end{figure}

The phase transition mentioned in the previous section occurs
for a fractional quantum Hall edge state such as $\nu=1/3$.
In the rest of this paper, we discuss another possibility for the phase transition
by considering a noisy gate voltage connected to an ohmic bath.
Such a situation can be realized, e.g., by connecting a long $LC$ transmission line
in series with the gate voltage~\cite{lehur04}.
The Hamiltonian describing the Caldeira-Leggett bath
of an infinite number of harmonic oscillators reads
\cite{weiss08}
\begin{gather}
H_{B}=\sum_{j}\left[\frac{P_{j}{}^{2}}{2M_{j}}
+\frac{M_{j}\Omega_{j}{}^{2}}{2}\left(X_{j}
-\frac{\lambda_{j}}{M_{j}\Omega_{j}{}^{2}}Q\right)^{2}\right],\label{eq:bath}
\end{gather}
where $\lambda_{j}$ is the strength of interaction
between the $j$th oscillator and the dot charge $Q$.
$X_{j}, P_{j}, M_{j}, \Omega_{j}$ denote
the normal coordinate, momentum, mass, and frequency
of the $j$th oscillator.
For an ohmic bath with impedance $R_{B}$,
the spectral density of the environmental coupling satisfies
\begin{gather}
J(\omega)\equiv\frac{\pi}{2}\sum_{j}\frac{\lambda_{j}{}^{2}}{M_{j}\Omega_{j}}
\delta(\omega-\Omega_{j})=R_{B}\omega.
\end{gather}
Below, we see how dissipation influences the
behavior of charge relaxation resistance.
To this end, we derive the euclidian effective action
for the system consisting of the mesoscopic capacitor and the ohmic bath.
By integrating out the quadratic degrees of freedom except
$\phi(x_{1})$ and $\phi(x_{2})$ from the kinetic term (\ref{eq:kinetic}),
we first obtain the effective action for the mesoscopic capacitor
\begin{align}
S_{MC}&=\frac{1}{\pi\nu\beta}\sum_{\omega_{n}}
\frac{|\omega_{n}|}{1-\rme^{-|\omega_{n}|L/v}}
|\tilde\phi(\omega_{n})|^{2}
+V\int\rmd\tau\cos[2\phi(\tau)]
+\int\rmd\tau\left[
\frac{E_{C}}{\pi^{2}}\{\phi(\tau)\}^{2}
-\frac{eV_{g}}{\pi}\phi(\tau)\right].\label{eq:action_mc}
\end{align}
Here $\phi\equiv[\phi(x_{1})-\phi(x_{2})]/2$,
$\tilde\phi(\omega_{n})$ is the Fourier transform of $\phi(\tau)$,
and $E_{C}\equiv e^{2}/(2C)$ is the charging energy.
Similarly, we can derive the effective action for the bath
from Eq.~(\ref{eq:bath}) as
\begin{gather}
S_{B}=\frac{\alpha}{\pi\beta}\sum_{\omega_{n}}|\omega_{n}||\tilde\phi(\omega_{n})|^{2}
\qquad\left(\alpha=\frac{R_{B}}{h/e^{2}}\right),\label{eq:action_e}
\end{gather}
where $\alpha$ is the strength of dissipation.
One can expect from Eqs.~(\ref{eq:action_mc}) and (\ref{eq:action_e})
that the coupling to the ohmic bath renormalizes the filling factor $\nu$,
which influences the critical point.
To discuss the quantum phase transition of the mesoscopic capacitor,
let us focus on the degeneracy point in the large $V$ region.
In this case, electron tunneling occurs between the dot and the reservoir
with small tunneling strength $t$.
Following Ref.~\cite{hamamoto10},
we can straightforwardly identify the scaling equations at low frequencies
$|\omega_{n}|\ll v/L$:
\begin{gather}
\dd{t}{l}=\left[1-\left(\alpha+\frac{1}{2\nu}\right)s^{2}\right]t,\qquad
\dd{s^{2}}{l}=-4s^{2}t^{2}\qquad(0<s\lesssim 1),
\end{gather}
which describes the Kosterlitz-Thouless transition.
One can see from these equations that, for $\alpha<1-1/(2\nu)$,
the tunneling strength always grows upon decreasing temperature.
For $\alpha>1-1/(2\nu)$, on the other hand,
it is possible that for sufficiently large $V$ (small $t$) that
the system is renormalized to a weak coupling configuration
with specified charge.
It should be noted that, unlike in Ref.~\cite{hamamoto10},
the critical phase transition occurs
even in the case of an integer quantum Hall edge ($\nu=1$)
when the bath impedance exceeds half a resistance quantum,
$R_{B}>h/(2e^{2})$.
Otherwise, the system flows towards the Kondo fixed point as $T\rightarrow0$,
and the charge relaxation resistance is given by the value for $V\rightarrow0$
as $R_{q}=h/(2e^{2})+R_{B}$, i.e., the noisy gate voltage
results in the shift of the quantized value by the bath impedance $R_{B}$.
The phase diagram of the universal quantization of the charge relaxation resistance
is shown in Fig.~\ref{fig:phase}.

\section{Summary}
We have theoretically studied the phase transition of dynamical resistance
in a mesoscopic capacitor applied a perpendicular magnetic field. 
For weak quasiparticle tunneling, 
the electrochemical capacitance can be expressed as a series combination of two capacitors,
one of which shows a behavior qualitatively different from the result of the scattering theory. 
We have also discussed the phase transition driven by
noise with with dissipation strength $\alpha$.
When $\alpha<1-1/(2\nu)$, the charge relaxation resistance is universally quantized
at $h/(2e^{2}\nu)+R_{B}$, with the bath impedance $R_{B}$.

\section*{Acknowledgements}
Y.H. and T.K. are grateful to T. Fujii for valuable discussions.
This research
was partially supported by JSPS and MAE under the
Japan-France Integrated Action Program (SAKURA) and by
a Grant-in-Aid for Young Scientists B No. 21740220
from the Ministry of Education, Science, Sports and Culture.
It was also supported by ANR-PNANO Contract MolSpin-
Tronics, No. ANR-06-NANO-27.

\end{document}